\documentclass[a4paper,11pt]{article}
\pdfoutput=1

\usepackage{jheppub}
\usepackage[T1]{fontenc}    % use 8-bit T1 fonts
\usepackage[dvipsnames]{xcolor}

\title{\boldmath Variational Autoencoders for New Physics Mining at the Large Hadron Collider}

\author[a]{Olmo Cerri,}
\author[a]{Thong Q. Nguyen,}
\author[b]{Maurizio Pierini,}
\author[a]{Maria Spiropulu}
\author[a]{and Jean-Roch Vlimant}

\affiliation[a]{California Institute of Technology,\\
                1200 E California Blvd, Pasadena, CA 91125, USA.}
\affiliation[b]{CERN,\\
                Espl. des Particules 1, 1217 Meyrin, Switzerland}
\emailAdd{olmo@caltech.edu}
\emailAdd{thong@caltech.edu}
\emailAdd{Maurizio.Pierini@cern.ch}
\emailAdd{smaria@caltech.edu}
\emailAdd{jvlimant@caltech.edu}

\abstract{
Using variational autoencoders trained on known physics processes, we develop a one-sided threshold test to isolate previously unseen processes as outlier events.  Since the autoencoder training does not depend on any specific new physics signature,
the proposed procedure doesn't make specific assumptions on the nature of new physics. An event selection based on this algorithm would be complementary to classic LHC searches, typically based on model-dependent hypothesis testing. Such an algorithm would deliver a list of anomalous events, that the experimental collaborations could further scrutinize and even release as a catalog, similarly to what is typically done in other scientific domains. Event topologies repeating in this dataset could inspire new-physics model building and new experimental searches. Running in the trigger system of the LHC experiments, such an application could identify anomalous events that would be otherwise lost, extending the scientific reach of the LHC.
}

\arxivnumber{1811.10276}

\begin{document}
\maketitle
\flushbottom

\section{Introduction}
\label{sec:intro}
One of the main motivations behind the construction of the CERN Large Hadron Collider (LHC) is the exploration of the high-energy frontier in search for {\it new physics} phenomena. New physics could answer some of the standing fundamental questions in particle physics, e.g., the nature of dark matter or the origin of electroweak symmetry breaking. In LHC experiments, searches for physics beyond the Standard Model (BSM) are typically carried on as fully-supervised data analyses: assuming a new physics scenario of some kind, a search is structured as a hypothesis test, based on a profiled-likelihood ratio~\cite{ATLAS:2011tau}. These searches are said to be {\it model dependent}, since they depend on considering a specific new physics model.

Assuming that one is testing the {\it right} model, this approach is very effective in discovering a signal, as demonstrated by the discovery of the Standard Model (SM) Higgs boson~\cite{Aad:2012tfa, Chatrchyan:2012xdj} at the LHC. On the other hand, given the (so far) negative outcome of many BSM searches at particle-physics experiments, it is possible that a future BSM model, if any, is not among those typically tested. The problem is more profound if analyzed in the context of the LHC big-data problem: at the LHC, 40 million proton-beam collisions are produced every second, but only $\sim$1000 collision events/sec can be stored by the ATLAS and CMS experiments, due to limited bandwidth, processing, and storage resources. It is possible to imagine BSM scenarios that would escape detection, simply because the corresponding new physics events would be rejected by a typical set of online selection algorithms.

Establishing alternative search methodologies with reduced model dependence is an important aspect of future LHC runs. Traditionally, this issue was addressed with so-called model-independent searches, performed at the Tevatron~\cite{Aaltonen:2008vt,Abazov:2011ma}, at HERA~\cite{Aaron:2008aa}, and at the LHC~\cite{CMS-PAS-EXO-14-016,Aaboud:2018ufy}, as discussed in Section~\ref{sec:relatedWorks}. 

In this paper, we propose to address this need by deploying an unsupervised algorithm in the online selection system (trigger) of the LHC experiments.\footnote{A description of the ATLAS and CMS trigger systems can be found in Ref.~\cite{Aaboud:2016leb} and Ref.~\cite{Khachatryan:2016bia}, respectively. In this study, we take the data-taking strategy of these two experiments as a reference. On the other hand, the proposed strategy could be adapted to other use cases.} This algorithm would be trained on known SM processes and could be able to identify BSM events as anomalies. The selected events could be stored in a special stream, scrutinized by experts (e.g., to exclude the occurrence of detector malfunctions that could explain the anomalies), and even released outside the experimental collaborations, in the form of an open-access catalog. The final goal of this application is to identify anomalous event topologies and inspire future supervised searches on data collected afterwards.

As an example, we consider the case of a typical single-lepton data stream, selected by a hardware-based Level-1 (L1) trigger system. In normal conditions, the L1 trigger is the first of a two-steps selection stage. After a coarse (and often local) reconstruction and loose selection at L1, events are fully reconstructed in the High Lever Trigger (HLT), where a much tighter selection is applied. The selection is usually done having in mind specific signal topologies, eg., specific BSM models. In this study, we imagine to replace this model-dependent selection with a variational autoencoder (VAE)~\cite{kingma2014auto,an2015variational} looking for anomalous events in the incoming single-lepton stream. The VAE is trained to compress the input event representation into a lower-dimension latent space and then decompress it, returning the shape parameters describing the  probability density function (pdf) of each input quantity given a point in the compressed space. In addition, a VAE allows a stochastic modeling of the latent space, a feature which is  missing in a simple AE architecture. The highlighted procedure is not specific of the considered single-lepton stream and could be easily extended to other data streams.

The distribution of the VAE's reconstruction loss on a validation sample is used to define a threshold, corresponding to a desired acceptance rate for SM events. All the events with loss larger than the threshold are considered as potential anomalies and could be stored in a low-rate anomalous-event data stream. In this work, we set the threshold such that $\sim 1000$ SM events would be collected every month under typical LHC operation conditions. In particular, we took as a reference 8 months of data taking per year, with an  integrated luminosity of $\sim40$~fb$^{-1}$. Assuming an LHC duty cycle of 2/3, this corresponds to an average instantaneous luminosity of $\sim 2.9\times 10^{33}$~cm$^{-2}$~s$^{-1}$.

We then evaluate the BSM production cross section that would correspond to a signal excess of 100 BSM events selected per month, as well as the one that would give a signal yield $\sim 1/3$ of the SM yield. For this, we consider a set of low-mass BSM resonances, decaying to one or more leptons and light enough to be challenging for the currently employed LHC trigger algorithms.  

This paper is structured as follows: we discuss related works in Section~\ref{sec:relatedWorks}. Section~\ref{sec:data} gives a brief description of the dataset used. Section~\ref{sec:unsupervised} describes the VAE model used in the study, as well as a set of fully-supervised classifiers used for performance comparison. Results are discussed in Section~\ref{sec:results}. In Section~\ref{sec:foreseen_applications} we discuss how such a procedure could be deployed in a typical LHC experiment while relying exclusively on data. Conclusions are given in Section~\ref{sec:conclusions}. Appendix~\ref{app:ae} provides a brief comparison between VAEs and plain autoencoders (AEs).

\section{Related work}
\label{sec:relatedWorks}
Model-independent searches for new physics have been performed at the Tevatron~\cite{Aaltonen:2008vt,Abazov:2011ma}, HERA~\cite{Aaron:2008aa}, and the LHC~\cite{CMS-PAS-EXO-14-016,Aaboud:2018ufy}. These searches are based on the comparison of a large set of binned distributions to the prediction from Monte Carlo (MC) simulations, in search for bins exhibiting a deviation larger than some predefined threshold. 
While the effectiveness of this strategy in establishing a discovery has been a matter of discussion, a recent study by the ATLAS collaboration~\cite{Aaboud:2018ufy} rephrased this model-independent search strategy into a tool to identify interesting excesses, on which traditional analysis techniques could be performed on independent datasets (e.g., the data collected after running the model-independent analysis). This change of scope has the advantage of reducing the trial factor (i.e., the so-called {\it look-elsewhere} effect~\cite{2008arXiv0811.1663L,Gross:2010qma}), which would otherwise wash out the significance of an observed excess. 

Our strategy is similar to what is proposed in Ref.~\cite{Aaboud:2018ufy}, with two substantial differences: (i) we aim to process also those events that could be discarded by the online selection, by running the algorithm as part of the trigger process; (ii) we do so exploiting  deep-learning-based anomaly detection techniques. 

Applying deep learning at the trigger level has been proposed in Ref.~\cite{TOPCLASS}.
Recent works~\cite{DAgnolo:2018cun,Collins:2018epr,DeSimone:2018efk,Hajer:2018kqm} have investigated the use of machine-learning techniques to setup new strategies for BSM searches with minimal or no assumption on the specific new-physics scenario under investigation. In this work, we use VAEs~\cite{kingma2014auto,an2015variational} based on high-level features as a baseline. Previously, autoencoders have been used in collider physics for detector monitoring~\cite{CMSdqm,CMSdc} and event generation~\cite{ATL-SOFT-PUB-2018-001}. Autoencoders have also been explored to define a jet tagger that would identify new physics events with anomalous jets~\cite{Heimel:2018mkt,AEjets}, with a strategy similar to what we apply to the full event in this work.

Anomaly detection has been a traditional use case for one-class machine learning methods, such as  one-class Support Vector Machine~\cite{scholkopf2001estimating} or Isolation Forest~\cite{liu2008isolation,liu2012isolation}. A review of proposed methods can be found in Ref.~\cite{aggarwal2015outlier}. 
Variational methods have been shown to be effective for novelty detection, as for instance is discussed in Ref.~\cite{deepmind}. In particular, VAEs~\cite{kingma2014auto} have been proposed as an effective method for anomaly detection~\cite{an2015variational}.

%One interesting feature that we explore in this work is the use of VAMP priors~\cite{VAMP} for optimal performance. To our knowledge, this is the first attempt to use this technique in particle physics. 

%Besides the baseline mode, we investigate the usage of variational autoencoders with a GRU unit, in order to start the analysis directly from a list of particles. In this context, we investigate potential improvement from VAMP priors, attention mechanism, and teacher 

\section{Data samples}
\label{sec:data}

The dataset used for this study is a refined version of the high-level-feature (HLF) dataset used in Ref.~\cite{TOPCLASS}. %Of the four data representations used there, we consider the list of  high-level features (HLF) and the list of reconstructed particles. Both representations are refined with respect to Ref.~\cite{TOPCLASS}, as described below. 
Proton-proton collisions are generated using the {\tt PYTHIA8} event-generation library~\cite{pythia}, fixing the center-of-mass energy to the LHC Run-II value (13~TeV) and the average number of overlapping collisions per beam crossing (pileup) to $20$. These beam conditions loosely correspond to the LHC operating conditions in 2016.

Events generated by {\tt PYTHIA8} are processed with the {\tt DELPHES} library~\cite{delphes}, to emulate detector efficiency and resolution effects. We take as a benchmark detector description the upgraded design of the CMS detector, foreseen for the High-Luminosity LHC phase~\cite{CMS_TP}. In particular, we use the CMS HL-LHC detector card distributed with {\tt DELPHES}.  We run the {\tt DELPHES} {\it particle-flow} (PF) algorithm, which combines the information from different detector components to derive a list of reconstructed particles, the so-called PF candidates. For each particle, the algorithm returns the measured energy and flight direction. Each particle is associated to one of three classes: charged particles, photons, and neutral hadrons. In addition, lists of reconstructed electrons and muons are given.

Many SM processes would contribute to the considered single-lepton dataset. For simplicity, we restrict the list of relevant SM processes to the four with the highest production cross sections, namely:
\begin{itemize}
\item Inclusive $W$ production, with $W\to \ell \nu$ ($\ell=e,~\mu,~\tau$).
\item Inclusive $Z$ production, with $Z\to \ell \ell$ ($\ell=e,~\mu,~\tau$).
\item $t \bar t$ production. 
\item QCD multijet production.\footnote{To speed up the generation process for QCD events, we require $\sqrt{\hat{s}}>10$~GeV, the fraction of QCD events with  $\sqrt{\hat{s}}<10$~GeV and producing a lepton within acceptance being negligible but computationally expensive.}
\end{itemize}
These samples are mixed to provide a SM cocktail dataset, which is then used to train autoencoder models and to tune the threshold requirement that defines what we consider an anomaly. The cocktail is built scaling down the high-statistics samples ($t \bar t$, $W$, and $Z$) to the lowest-statistics one (QCD, whose generation is the most computing-expensive), according to their production cross-section values (estimated at leading order with {\tt PYTHIA}) and selection efficiencies, shown in Table~\ref{tab:eff}.

\begin{table}[!htb]
  \caption{Acceptance and L1 trigger (i.e. $p_T^\ell$ and $\textsc{Iso}$ requirement) efficiency for the four studied SM processes and corresponding values for the BSM benchmark models. For SM processes, we quote the total cross section before the trigger, the expected number of events per month and the fraction in the SM cocktail. For BSM models, we compute the production cross section corresponding to an average of 100 BSM events per month passing the acceptance and L1 trigger requirements. The monthly event yield is computed assuming an average luminosity per month of $5 \text{ fb}^{-1}$, corresponding to the running conditions discussed in Section~\ref{sec:intro}.\label{tab:eff}}
  
\vspace{10pt}

\centering
\begin{tabular}{c|ccc|cc}
\hline
\multicolumn{6}{c}{Standard Model processes} \\
\hline
Process    & Acceptance & L1 trigger & Cross & Event & Events \\
           & & efficiency & section [nb] & fraction & /month \\
\hline
$W$        & $55.6\%$ & $68\%$ & $58$ & $59.2\%$ & 110M \\
QCD        & $0.08\%$ & $9.6\%$ & $1.6 \cdot 10^5$ & $33.8\%$ & 63M \\
$Z$        & $16\%$ & $77\%$ & $20$ & $6.7\%$ & 12M \\
$t \bar t$ & $37\%$ & $49\%$ & $0.7$ & $0.3\%$ & 0.6M \\
\hline
\end{tabular}
\begin{tabular}{c|cc|cc}
\multicolumn{5}{c}{BSM benchmark processes} \\
\hline
Process        &  Acceptance & L1 trigger & Total & Cross-section\\
               &  & efficiency & efficiency & 100 BSM events/month\\
\hline
$A \to 4\ell$  & $5\%$ & $98\%$ & $5\%$ & 0.44~pb\\
$LQ\to b \tau$ & $19\%$ & $62\%$ & $12\%$ & 0.17~pb\\ 
$h^0\to\tau\tau$ & $9\%$ & $70\%$ & $6\%$ & 0.34~pb\\
$h^{\pm}\to\tau\nu$     & $18\%$ & $69\%$ & $12\%$ & 0.16~pb\\
\hline
\end{tabular}
\end{table}

Events are filtered at generation requiring an electron, muon, or tau lepton with $p_T>22$~GeV. Once detector effects are taken into account through the {\tt DELPHES} simulation, events are further selected requiring the presence of one reconstructed lepton (electron or muon) with transverse momentum $p_T>23$~GeV and a loose isolation requirement $\textsc{Iso}<0.45$. If more than one reconstructed lepton is present, the highest $p_T$ one is considered.
The isolation for the considered lepton $\ell$ is computed as:
\begin{equation}
\textsc{Iso} = \frac{\sum_{p \neq \ell} p_T^p}{p_T^\ell}~,
\end{equation}
where the index $p$ runs over all the photons, charged particles, and neutral hadrons within a cone of size $\Delta R = \sqrt{\Delta \eta^2 + \Delta \phi^2}<0.3$ from $\ell$.\footnote{As common for collider physics, we use a Cartesian coordinate system with the $z$ axis oriented along the beam axis, the $x$ axis on the horizontal plane, and the $y$ axis oriented upward. The $x$ and $y$ axes define the transverse plane, while the $z$ axis identifies the longitudinal direction. The azimuth angle $\phi$ is computed from the $x$ axis. The polar angle $\theta$ is used to compute the pseudorapidity $\eta = -\log(\tan(\theta/2))$. We fix units such that $c=\hbar=1$.} 

The 21 considered HLF quantities are:
\begin{itemize}
\item The absolute value of the isolated-lepton transverse momentum $p_T^\ell$.
\item The three isolation quantities (\textsc{ChPFIso}, \textsc{NeuPFIso}, \textsc{GammaPFIso}) for the isolated lepton, computed with respect to charged particles, neutral hadrons and photons, respectively.
\item The lepton charge.
\item A Boolean flag (\textsc{isEle}) set to 1 when the trigger lepton is an electron, 0 otherwise.
\item $S_T$, i.e. the scalar sum of the $p_T$ of all the jets, leptons, and photons in the event with $p_T>30$~GeV and $|\eta|<2.6$. Jets are clustered from the reconstructed PF candidates, using the {\tt FASTJET}~\cite{fastjet} implementation of the anti-$k_T$ jet algorithm~\cite{antikt}, with a jet-size parameter R=0.4.
\item The number of jets entering the $S_T$ sum ($N_J$).
\item The invariant mass of the set of jets entering the $S_T$ sum ($M_J$).
\item The number of these jets being identified as originating from a $b$ quark ($N_b$).
\item The missing transverse momentum, decomposed into its parallel ($p_{T,\parallel}^{\text{miss}}$) and orthogonal ($p_{T,\perp}^{\text{miss}}$) components with respect to the lepton $\ell$ direction. The missing transverse momentum is defined as the negative sum of the PF-candidate $p_T$ vectors:
\begin{equation}
\vec{p}_T^{\text{~miss}} = -\sum_{q} \vec{p}_T^{~q}~.
\end{equation}
\item The transverse mass, $M_T$, of the isolated lepton $\ell$ and the $\vec{p}_T^{\text{~miss}}$ system, defined as:
\begin{equation}
M_T = \sqrt{2p_T^{\ell}E_T^{\text{miss}}(1-\cos{\Delta \phi})}~,
\end{equation}
with $\Delta \phi$ the azimuth separation between the $\vec{p}_T^{~\ell}$ and $\vec{p}_T^{\text{~miss}}$ vectors, and $E_T^{\text{miss}}$ the magnitude of $\vec{p}_T^{\text{~miss}}$.
\item The number of selected muons ($N_\mu$).
\item The invariant mass of this set of muons ($M_\mu$). 
\item The absolute value of the total transverse momentum of these muons ($p_{T,TOT}^{\mu}$).
\item The number of selected electrons ($N_e$).
\item The invariant mass of this set of electrons ($M_e$). 
\item The absolute value of the total transverse momentum of these electrons ($p_{T,TOT}^{e}$).
\item The number of reconstructed charged hadrons.
\item The number of reconstructed neutral hadrons.
\end{itemize}
This list of HLF quantities is not defined having in mind a specific BSM scenario. Instead, it is conceived to include relevant information to discriminate the various SM processes populating the single-lepton data stream. On the other hand, it is generic enough to allow (at least in principle) the identification of a large set of new physics scenarios. 

In addition to the four SM processes listed above, we consider the following BSM models to benchmark anomaly-detection capabilities:
\begin{itemize}
\item A leptoquark $LQ$ with mass 80~GeV, decaying to a $b$ quark and a $\tau$ lepton. 
\item A neutral scalar boson with mass 50~GeV, decaying to two off-shell $Z$ bosons, each forced to decay to two leptons: $A \to 4\ell$.
\item A scalar boson with mass 60~GeV, decaying to two tau leptons: $h^0\to \tau \tau$.
\item  A charged scalar boson with mass 60~GeV, decaying to a tau lepton and a neutrino: $h^\pm \to \tau \nu$.
\end{itemize}
For each BSM scenario, we consider any direct production mechanism implemented in ${\tt PYTHIA8}$, including associate jet production. We list in Table~\ref{tab:eff} the leading-order production cross section and selection efficiency for each model.

\begin{figure}[!htb]
    \centering
    \includegraphics[width=\textwidth]{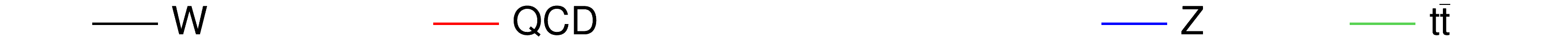}

    \includegraphics[width=\textwidth]{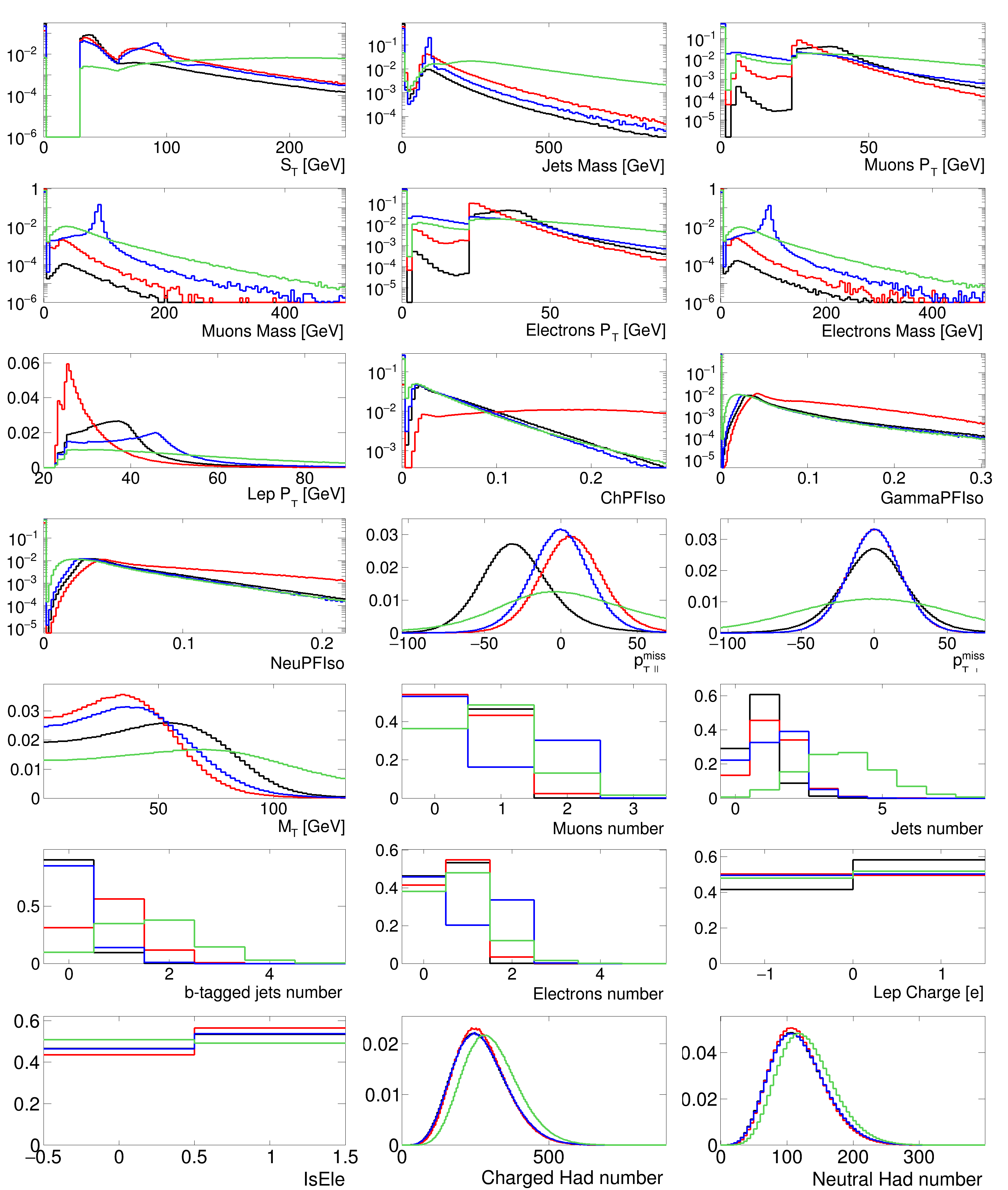}
    
    %\vspace{0.3cm}
    
    \caption{Distribution of the HLF quantities for the four considered SM processes.\label{fig:SM_HLF}}
\end{figure}

\begin{figure}[tb]
    \centering
    \includegraphics[width=\textwidth]{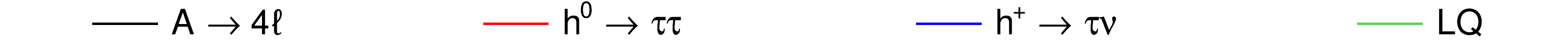}

    \includegraphics[width=\textwidth]{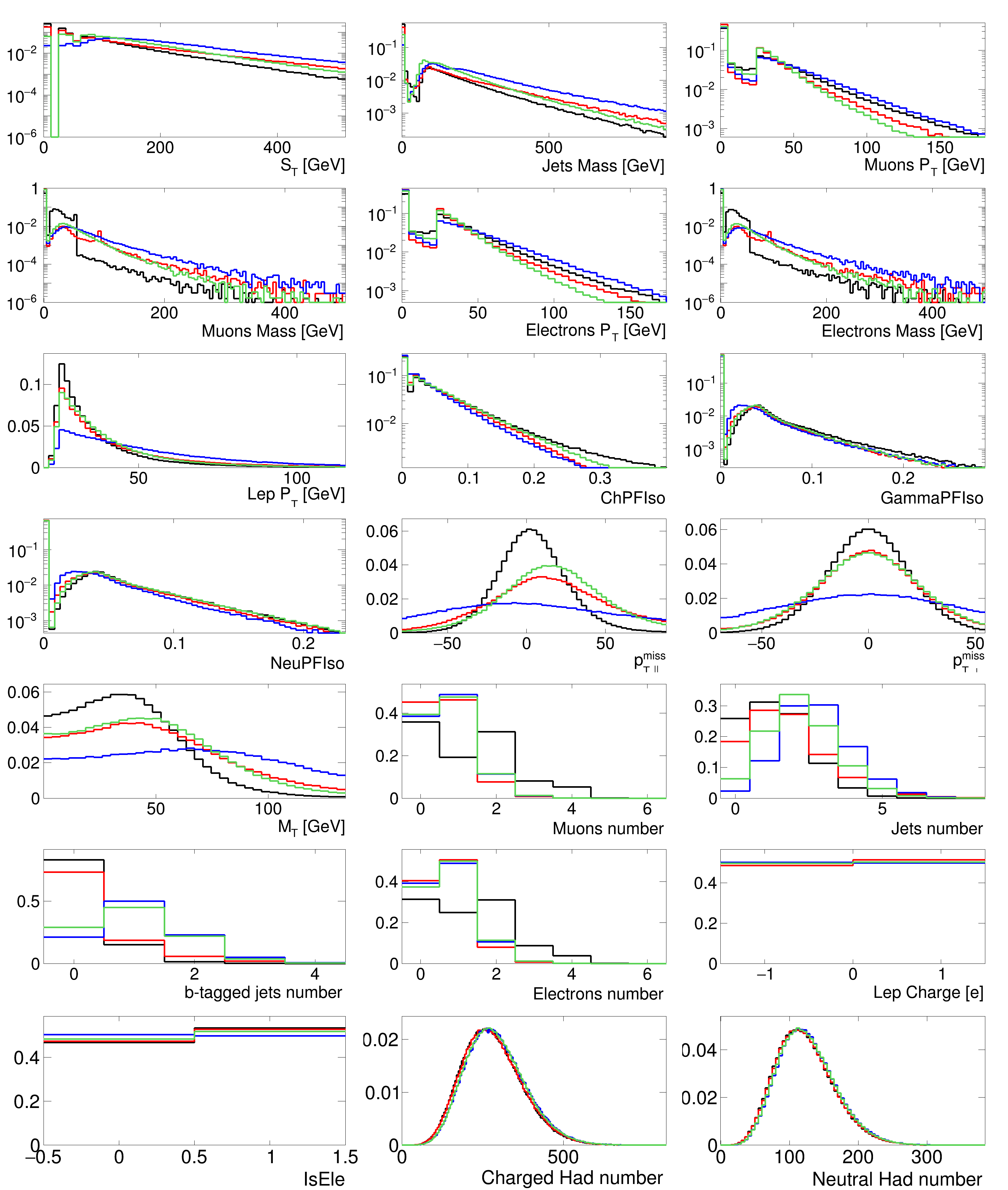}
    
    \vspace{0.3cm}
    
    \caption{Distribution of the HLF quantities for the four considered BSM benchmark models.\label{fig:BSM_HLF}}
\end{figure}

Figures~\ref{fig:SM_HLF}~and~\ref{fig:BSM_HLF} show the distribution of HLF quantities for the SM processes and the BSM benchmark models, respectively.

\section{Model description}
\label{sec:unsupervised}

We train VAEs on the SM cocktail sample described in Section~\ref{sec:data}, taking as input the 21 HLF quantities listed there. The use of HLF quantities to represent events limits the model independence of the anomaly detection procedure. While the list of features is chosen to represent the main physics aspects of the considered SM processes and is in no way tailored to specific BSM models, it is true that such a list might be more suitable for certain models than for others. In this respect, one cannot guarantee that the anomaly-detection performance observed on a given BSM model would generalize to any BSM scenario. We will address in a future work a possible solution to reduce the residual model dependence implied by the input event representation.

In this section, we present both the best-performing autoencoder model, trained to encode and decode the SM training sample, and a set of four supervised classifiers, each trained to distinguish one of the four BSM benchmark models from SM events. We use the classification performance of these supervised algorithms as an estimate of the best performance that the VAE could get to.

\subsection{Autoencoders}
Autoencoders are algorithms that compress a given set of inputs variables in a latent space (encoding) and then, starting from the latent space, reconstruct the HLF input values (decoding). The loss distribution of an AE is used in the context of anomaly detection to isolate potential anomalies. Since the compression capability learned on a given sample doesn't typically generalize to other samples, the tails of the loss distribution could be enriched by new kinds of events, different than those used to train the model. In the specific case considered in this study, the tail of the loss distribution for an AE trained on SM data might be enriched with BSM events.

In this work we focus on VAEs~\cite{kingma2014auto}. For each event, a plain AE predicts an encoded point in the latent space and a decoded point in the original space. In other words, AEs are point-estimate algorithms. VAEs, instead, associate to each input event an estimated probability distributions in the latent space and in the original space. Doing so, VAEs provide both a best-point estimate and an estimate of the associated statistical noise. Besides this conceptual difference, VAEs have been shown to provide competitive performances for novelty~\cite{deepmind} and anomaly~\cite{an2015variational} detection.

\begin{figure}[!htbp]
    \centering
    \includegraphics[width=0.76\textwidth]{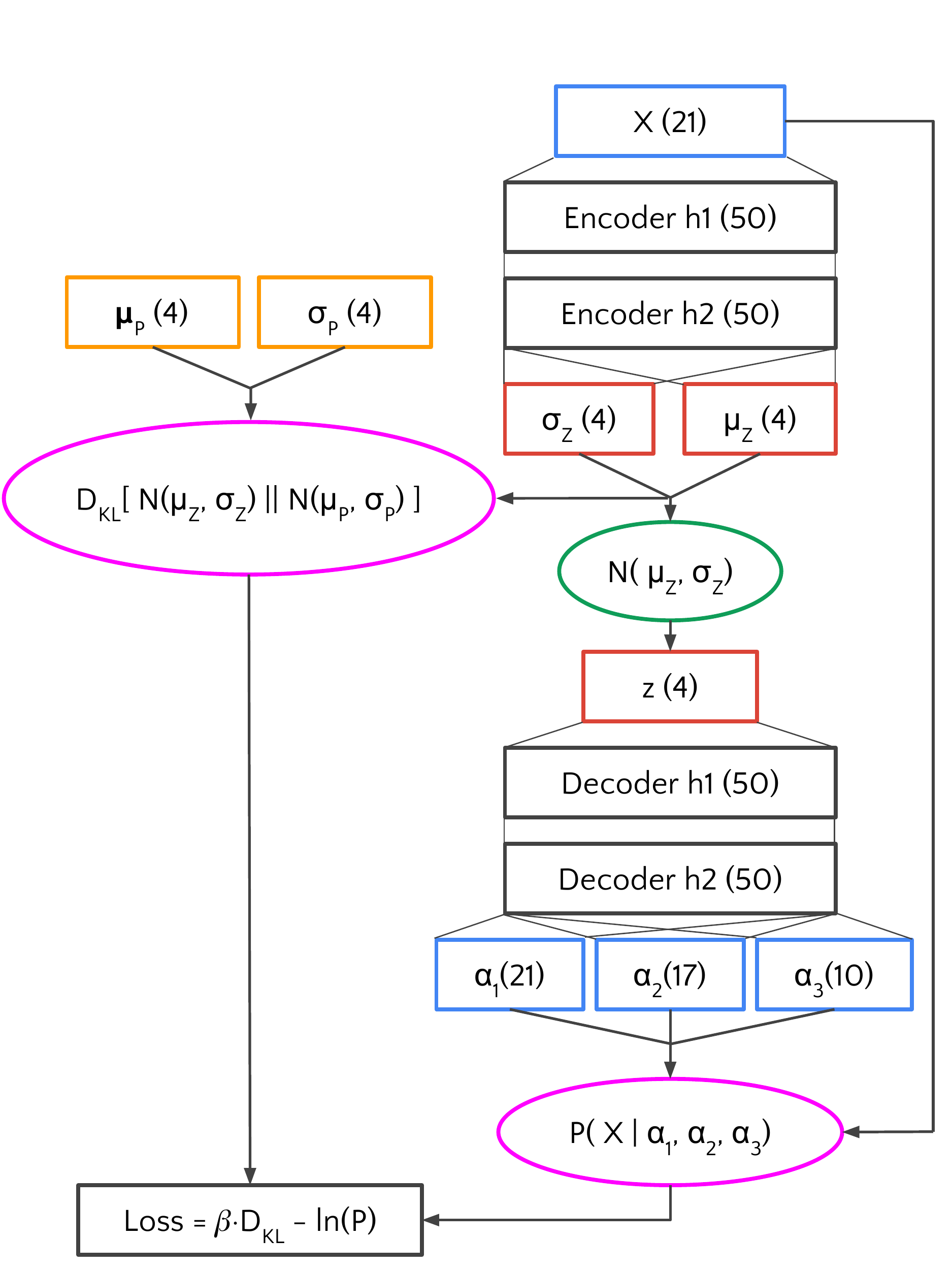}
    \caption{Schematic representation of the VAE architecture presented in the text. The size of each layer is indicated by the value within brackets. The \textcolor{RoyalBlue}{blue rectangle $X$} represents the input layer, which is connected to a stack of two consecutive fully connected layers (black boxes). The last of the two black box is connected to two layers with four nodes each (\textcolor{Red}{red boxes}), representing the $\mu_z$ and $\sigma_z$ parameters of the encoder pdf $p(z|x)$. The \textcolor{Green}{green oval} represents the sampling operator, which returns a set of values for the 4-dimensional \textcolor{Red}{latent variables $z$}. These values are  fed into the decoder, consisting of two consecutive hidden layers of 50 nodes each (black boxes). The last of the decoder hidden layer is connected to the \textcolor{RoyalBlue}{three output layers}, whose nodes correspond to the parameters of the predicted distribution in the initial 21-dimension space. The \textcolor{Magenta}{pink ovals} represent the computation of the two parts of the loss function: the KL loss and the reconstruction loss (see text). The \textcolor{Magenta}{computation of the KL} requires 8 additional learnable parameters ($\mu_p$ and $\sigma_p$, represented by the \textcolor{RedOrange}{orange boxes} on the top-left part of the figure), corresponding to the means and RMS of the four-dimensional Gaussian prior $p(z)$. The total loss in computed as described by the formula in the bottom-left black box (see Eq.~(\ref{eq:loss})). \label{fig:VAE_schematics}}
\end{figure}

We consider the VAE architecture shown in Fig.~\ref{fig:VAE_schematics}, characterized by a four-dimensional latent space. Each latent dimension is associated to a Gaussian pdf and its two degrees of freedom (mean $\mu_z$ and RMS $\sigma_z$). The input layer 
consists of 21 nodes, corresponding to the 21 HLF quantities described in Section~\ref{sec:data}. This layer is connected to the latent space through a stack of two fully connected layers, each consisting of 50 nodes with ReLU activation functions. Two four-node layers are fully connected to the second 50-node layer. Linear activation functions are used for the first of these four-node layers, interpreted as the set of four $\mu_z$ of the four-dimension Gaussian pdf $p(z)$. The nodes of the second layer are activated by the functions:
\begin{equation}
  \displaystyle \text{p-ISRLu}(x) = 1 + 5\cdot10^{-3} + \Theta(x)x +\Theta(-x)\frac{x}{\sqrt{1+x^2}}~.
\end{equation}
This activation allows to improve the training stability, being strictly positive defined, non linear, and with no exponentially growing term (which might have created instabilities in the early epochs of the training). The four nodes of this layer are interpreted as the $\sigma_z$ parameters of $p(z)$. After several trials, the dimension of the latent space has been set to 4 in order to keep a good training stability without impacting the VAE performances.
The decoding step originates from a point in the latent space, sampled according to the predicted pdf (green oval in Fig.~\ref{fig:VAE_schematics}). The coordinates of this point in the latent space are fed into a sequence of two hidden dense layers, each consisting of 50 neurons with ReLU activation functions. The last of these layers is connected to three dense layers of 21, 17, and 10 neurons, activated by linear, p-ISRLu and clipped-tanh functions, respectively. The clipped-tanh function if written as:
\begin{equation}
\displaystyle C_{\tanh}(x) = \frac{1}{2}(1 + 0.999\cdot\tanh{x})~.
\end{equation}
Given the latent-space representation, the 48 output nodes represent the parameters of the pdfs describing the input HLF probability, i.e., the $\alpha$ parameters of Eq.(\ref{eq:Lossreco}).

The total VAE loss function $\text{Loss}_\text{Tot}$ is a weighted sum of two pieces~\cite{higgins2016beta}: a term related to the reconstruction likelihood ($\text{Loss}_\text{reco}$) and the Kullback-Leibler divergence ($D_\text{KL}$) between the latent space pdf and the prior:
\begin{equation}
\text{Loss}_\text{Tot} = \text{Loss}_\text{reco} + \beta D_\text{KL}~,
\label{eq:loss}
\end{equation}
where $\beta$ is a free parameter. We fix $\beta=0.3$, for which we obtained good reconstruction performances.\footnote{Following Ref.~\cite{higgins2016beta}, we tried to increase the value of $\beta$ up to 4 without observing a substantial difference in performance.} The prior $p(z)$ chosen for the latent space is a four-dimension Gaussian with a diagonal covariance matrix. The means ($\mu_P$) and the diagonal terms of the covariance matrix ($\sigma_P$) are free parameters of the algorithm and are optimized during the back-propagation. The Kullback-Leibler divergence between two Gaussian distributions has an analytic form. Hence, for each batch, $D_\text{KL}$ can be expressed as:
\begin{equation}
\begin{split}
    D_\text{KL} & = \frac{1}{k}\sum_i D_\text{KL}\left( N(\mu_z^{i}, \sigma_z^{i})\ \mid\mid\ N(\mu_P, \sigma_P)\right) \\
    & = \frac{1}{2k}\sum_{i,j} \left(\sigma_P^{j}\sigma_z^{i,j}\right)^2 + \left(\frac{\mu_P^j - \mu_z^{i,j}}{\sigma_P^j}\right)^2 + \ln
    \frac{\sigma_P^j}{\sigma_z^{i,j}} -1~,
\end{split}
\end{equation}
where $k$ is the batch size, $i$ runs over the samples and $j$ over the latent space dimensions. Similarly, $\text{Loss}_\text{reco}$ is the average negative-log-likelihood of the inputs given the predicted $\alpha$ values:
\begin{equation}
\begin{split}
    \text{Loss}_\text{reco} &= -\frac{1}{k} \sum_i \ln\left[P(x\mid \alpha_1, \alpha_2, \alpha_3)\right]\\
    &= -\frac{1}{k} \sum_{i,j} \ln\left[f_j(x_{i,j}\mid \alpha_1^{i,j}, \alpha_2^{i,j}, \alpha_3^{i,j})\right]~.
\end{split}
\label{eq:Lossreco}
\end{equation}
In the equation, $j$ runs over the input space dimensions, $f_j$ is the functional form chose to describe the pdf of the $j$-th input variable and $\alpha_m^{i,j}$ are the parameter of the function.
Different functional forms have been chosen for $f_j$, to properly describe  different classes of HLF distributions:
\begin{itemize}
    \item {\bf Clipped Log-normal + $\delta$ function}: used to describe  $S_T$, $M_J$, $p_T^\mu$, $M_\mu$, $p_T^e$, $M_e$, $p_T^\ell$, {\textsc ChPFIso}, {\textsc NeuPFIso} and {\textsc GammaPFIso}:
      \begin{equation}
            P(x \mid \alpha_1, \alpha_2, \alpha_3) = 
            \left\{
                \begin{array}{ll}
                    \alpha_3\delta(x) +     \frac{1-\alpha_3}{x\alpha_2\sqrt{2\pi}}\exp\left(-\frac{(\ln x - \alpha_1)^2}{2\alpha_2^2}\right)~for~x \geq 10^{-4} \\
                    0~for~x < 10^{-4}
                \end{array}
          \right.~.
        \end{equation}
    \item {\bf Gaussian}: used for $p_{T,\parallel}^{\text{miss}}$ and $p_{T,\perp}^{\text{miss}}$: 
     \begin{equation}
         P(x \mid \alpha_1, \alpha_2) = \frac{1}{\alpha_2\sqrt{2\pi}}\exp\left(-\frac{(x - \alpha_1)^2}{2\alpha_2^2}\right)~.
     \end{equation}
    \item {\bf Truncated Gaussian}: a Gaussian function truncated for negative values and normalized to unit area for $X>0$. Used to model $M_T$:
       \begin{equation}
      P(x \mid \alpha_1, \alpha_2) = \Theta(x)\cdot  \frac{1 + 0.5\cdot(1+ \text{erf}\frac{-\alpha_1}{\alpha_2\sqrt{2}})}{\alpha_2\sqrt{2\pi}}\exp\left(-\frac{(x - \alpha_1)^2}{2\alpha_2^2}\right)~.
       \end{equation}
    \item {\bf Discrete truncated Gaussian}: like the truncated Gaussian, but normalized to be evaluated on integers (i.e. $\displaystyle \sum_{n=0}^{\infty} P(n) = 1$). This function is used to describe $N_\mu$, $N_e$, $N_b$ and $N_J$. It is written as: 
    \begin{equation}
                P(n \mid \alpha_1, \alpha_2) = \Theta(x)
                \left[ \text{erf}\left(\frac{n+0.5-\alpha_1}{\alpha_2\sqrt{2}}\right)
                -
                \text{erf}\left(\frac{n-0.5-\alpha_1}{\alpha_2\sqrt{2}}\right)
                \right]
                \mathcal{N}~,
    \end{equation}
        where the normalization factor $\mathcal{N}$ is set to:
                 \begin{equation}
                \mathcal{N}=
                1+\frac{1}{2}\left(1+\text{erf}\left(\frac{-0.5 -\alpha_1}{\alpha_2\sqrt{2}}\right)\right).
    \end{equation}
    \item {\bf Binomial}: used for (\textsc{isEle}) and lepton charge:
    \begin{equation}
        P(n \mid p) = \delta_{n,m}p + \delta_{n,l}(1-p)~,
    \end{equation}
    where $m$ and $l$ are the two possible values of the variable (0 or 1 for (\textsc{isEle}) and -1 or 1 for lepton charge) and $p = C_{\tanh}(\alpha_1)$
    \item {\bf Poisson}: used for charged-particle and neutral-hadron multiplicities:
    \begin{equation}
        P(n \mid \mu) = \frac{\mu^n e^{-\mu}}{\Gamma(n+1)}~,
    \end{equation}
    where $\mu = \text{p-ISRLu}(\alpha_1)$.
\end{itemize}

\begin{figure}[!htb]
\centering
\includegraphics[width=\textwidth]{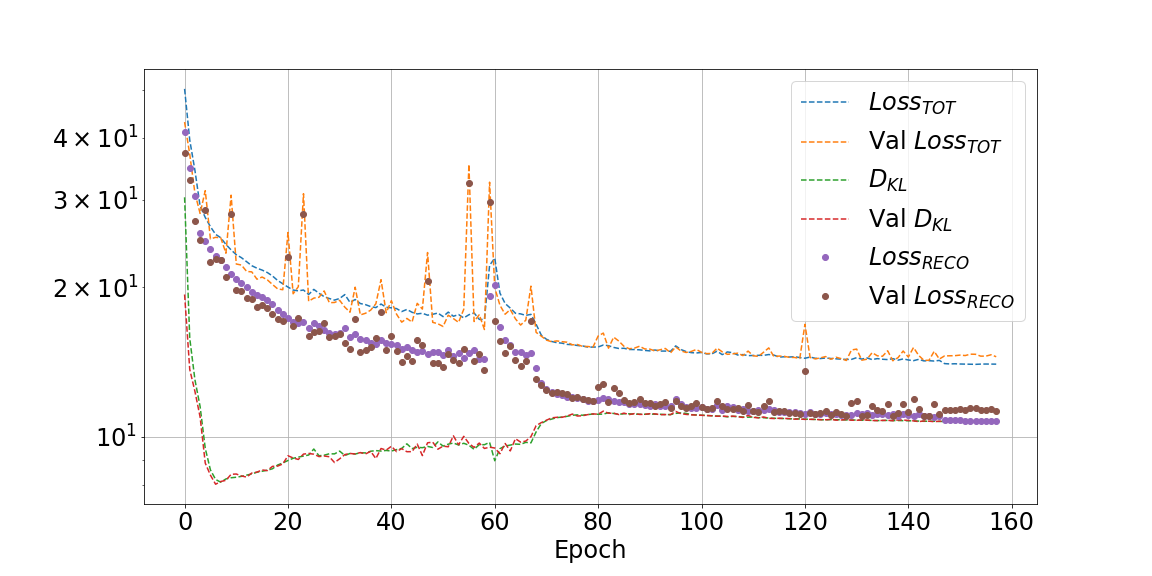}
\caption{Training history for VAE. Total loss, reconstruction negative-log-likelihood ($\text{Loss}_\text{reoc}$) and KL divergence ($D_{KL}$) are shown separately for training and validation set though all the training epochs.\label{fig:VAEtraining}}
\end{figure}

These custom functions provide an improved performance with respect to the standard choice of an MSE loss. When using the MSE loss, one is implicitly writing the likelihood of the input quantities as a product of Gaussian functions with equal variance. This choice is clearly a poor description of the input distributions at hand in this application and it results in a poor representation of the cores and the tails of the input distributions. Instead, the use of these tailored functions allows to correctly describe the distribution cores and to improve the description of the tails. 

We point out that the final performance depends on the choice of the p(x|z) functional form (i.e., on  the modeled dependence of the observed features on the latent variables) and the p(z) prior function. The former was tuned looking at the distributions for SM events. The latter is arbitrary. We explored techniques to optimize the choice of p(z), learning it from the data~\cite{VAMP}. In this case, no practical advantage in terms of anomaly detection was observed. An improved choice of  p(x|z) and the possibility of learning p(z) during the train could potentially further boost the performances of this algorithm and will be the subject of future studies with real LHC collision data.

The model shown in Fig.~\ref{fig:VAE_schematics} is  implemented in  {\tt KERAS+TENSORFLOW}~\cite{keras,tensorflow}, trained with the Adam optimizer~\cite{adam} on a SM dataset of 3.45M events, equivalent to an integrated luminosity of $\sim 100$~pb$^{-1}$. The SM validation dataset is made of 3.45M of statistically independent examples. Such a sample would be collected in about ten hours of continuous run, under the assumptions made in this study (see Section~\ref{sec:intro}). In training, we fix the batch size to 1000. We use early stopping with patience set to 20 and $\delta_{\mathrm{min}} =0.005$, and we progressively reduce the learning rate on plateau, with patience set to 8 and $\delta_{\mathrm{min}} =0.01$.
% As a cross check, the model was also implemented in {\tt PYTORCH}~\cite{pytorch}, with which consistent results were derived\textbf{Thong's results in pytorch are much worse}. 
%All the results showed below are derived with the {\tt KERAS+TENSORFLOW} implementation.

\begin{figure}[hp]
    \centering
    \includegraphics[width=0.87\textwidth]{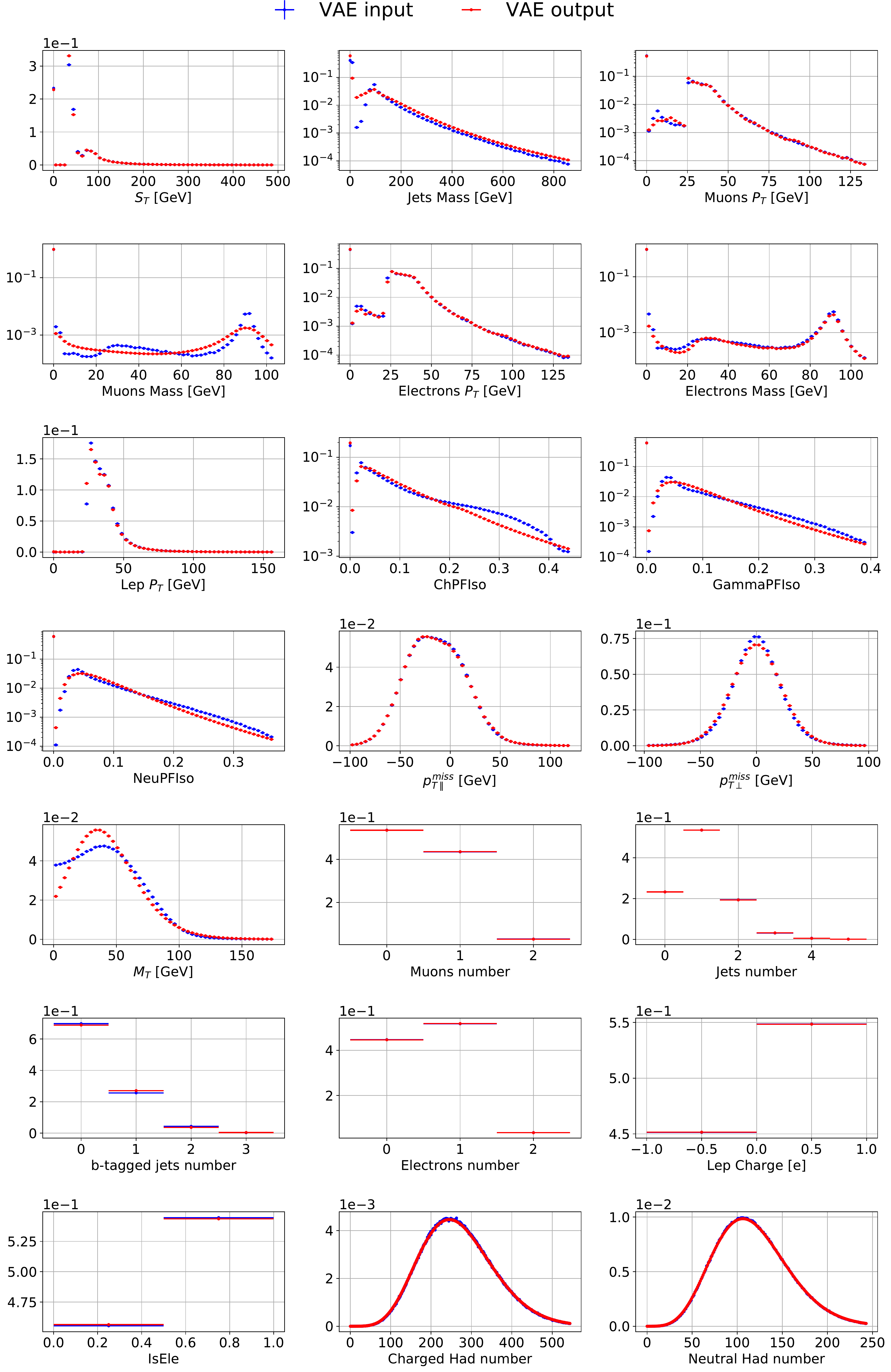}
    \vspace{-10pt}
    \caption{Comparison of input (blue) and output (red) probability distributions for the HLF quantities in the validation sample. The input distributions are normalized to unity. The output distributions are obtained summing over the predicted pdf of each event, normalized to the inverse of the total number of events (so that the total sum is normalized to unity).\label{fig:encoding-decoding}}
\end{figure}

The model's training history is shown in Fig.~\ref{fig:VAEtraining}. Figure~\ref{fig:encoding-decoding} shows the comparison of the input and output distributions for the 21 HLF quantities in the validation dataset. A general good agreement is observed on the bulk of the distributions, even if some of the distributions are not well described on the tails. These discrepancies don't have a sizable impact on the anomaly-detection strategy, as shown in Section.~\ref{sec:results}. Nevertheless, alternative architectures were tested, in order to reduce these discrepancies. For instance, we increased or decreased the dimensionality of the latent space, we changed the value of $\beta$ in Eq.(\ref{eq:loss}), we changed the number of neurons in the hidden layers, tried the RMSprop optimizer, and used plain Gaussian functions to describe the 21 input features. Some of these choices improved the encoding-decoding capability of the VAE, with up to a 10\% decrease of the loss function at the end of the training. On the other hand, none of these alternative models provided a sizable improvement in the anomaly-detection performance. For simplicity, we decided to limit our study to the architecture in Fig.~\ref{fig:VAE_schematics} and dropped these alternative models.

\subsection{Supervised classifiers}
\label{sec:supervised}
For each of the four BSM benchmark models, we train a fully-supervised classifier, based on a Boosted Decision Tree (BDT).
Each BDT receives as input the same 21 features used by the VAE and is trained on a labeled dataset consisting of the SM cocktail (the background) and one of the four BSM benchmark models (the signal). The implementation is done through the Gradient Boosted Classifier of the scikit-learn library~\cite{scikit-learn}. The algorithm was tuned  with up to 150 estimators, minimum samples per leaf and maximum depth equal to 3, a learning rate of 0.1, and a tolerance of $10^{-4}$ on the validation loss function (choose to be the default deviance). Each BDT, tailored to a specific BSM model, is trained on 3.45M SM events and about 0.5M BSM events, consistently up-weighted in order to match the size of the SM sample during the training.

\begin{table}[!htb]
\caption{Classification performance of the four BDT classifiers described in the text, each trained on one of the four BSM benchmark models. The two set of values correspond to the area under ROC curve (AUC), and to the true positive rate (TPR) for a SM false positive rate $\epsilon_{SM} = 5.4\cdot 10^{-6}$, i.e., to $\sim 1000$ SM events accepted every month.\label{tab:DNN_results}}

\vspace{10pt}

\centering
\begin{tabular}{c|cc}
\hline
Process & AUC & TPR [$\%$]\\
\hline
$A \to 4\ell$  & 0.98 & 5.4 \\
$LQ\to b \tau$ & 0.94 & 0.2 \\ 
$h^0 \to \tau\tau$ & 0.90 & 0.1 \\
$h^\pm \to \tau\nu$ & 0.97 & 0.3 \\
\hline
\end{tabular}
\end{table}

\begin{figure}[!htb]
\centering
\includegraphics[width=0.7\textwidth]{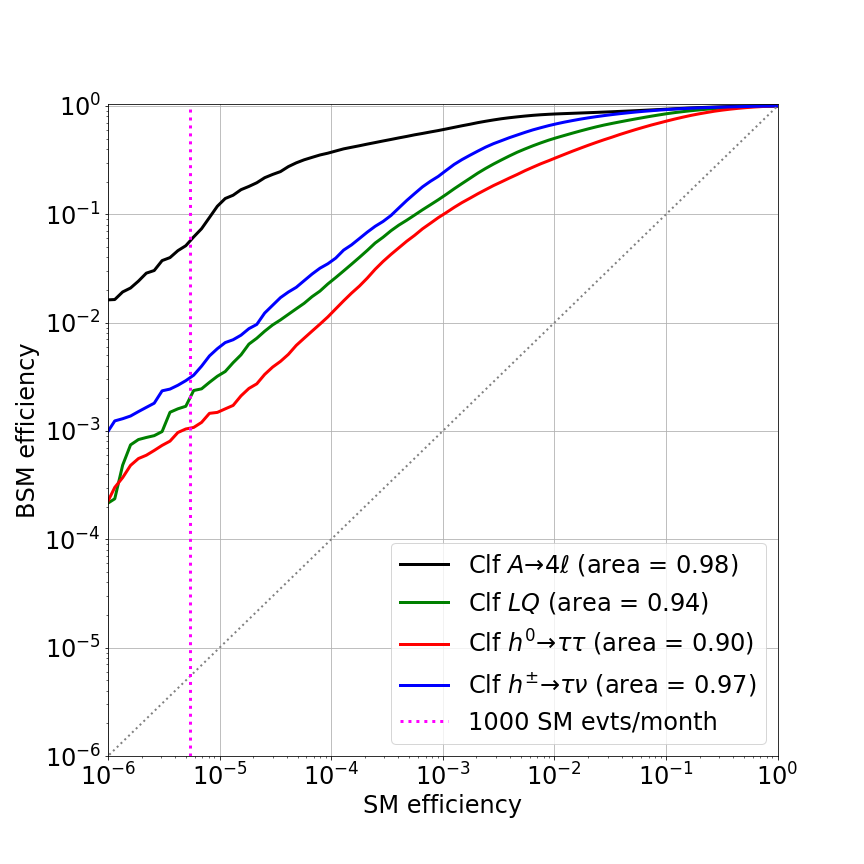}
\caption{ROC curves for the fully-supervised BDT classifiers, optimized to separate each of the four BSM benchmark models from the SM cocktail dataset.\label{fig:ROC_BDT}}
\end{figure}

We show in Table~\ref{tab:DNN_results} and in Figure~\ref{fig:ROC_BDT} the classification performance of the four supervised BDTs, which set a qualitative upper limit for VAE's results. Overall, the four models can be discriminated with good accuracy, with some loss of performance for those models sharing similarities with specific SM processes (e.g., $h^0 \to \tau \tau$ exhibiting single- and double-lepton topology with missing transverse energy, typical of $t \bar t$ events). In the table, we also quote the true-positive rate (TPR) for each BSM model corresponding to a working point of SM false positive rate $\epsilon_{SM} = 5.4\cdot 10^{-6}$, corresponding to an average of $\sim 1000$ SM events accepted every month.

In addition to BDTs, we experimented with fully-connected deep neural networks (DNNs) with two hidden layers. Despite trying different architectures, we didn't find a configuration in which the DNN classifiers could outperform the BDTs. This is due to the fact that, given the limited complexity of the problem at hand, a simple BDT can extract the maximum discrimination power from the 21 inputs. The limiting factor preventing to reach larger auc values is not to be found in the model complexity but in the discriminating power of the 21 input features. Not being tailored on the benchmark BSM scenarios, these features don't carry all the needed information for an optimal signal-to-background separation. While certainly one could obtain a better performance with more tailored classifiers, the purpose of this exercise was to provide a fair comparison for the VAE. In view of these considerations, we decided to use the BDTs as reference supervised classifiers.

\section{Results with VAE}
\label{sec:results}

An event is classified as anomalous whenever the associated loss, computed from the VAE output, is above a given threshold. Since no BSM signal has been observed by LHC experiments so far, it is reasonable to expect that a new-physics signal, if any, would be characterized by a low production cross section and/or features very similar to those of a SM process. In view of this, we decided to use a tight threshold value, in order to reduce as much as possible any SM contribution. 

\begin{figure}[!tb]
    \centering
    \includegraphics[width=\textwidth]{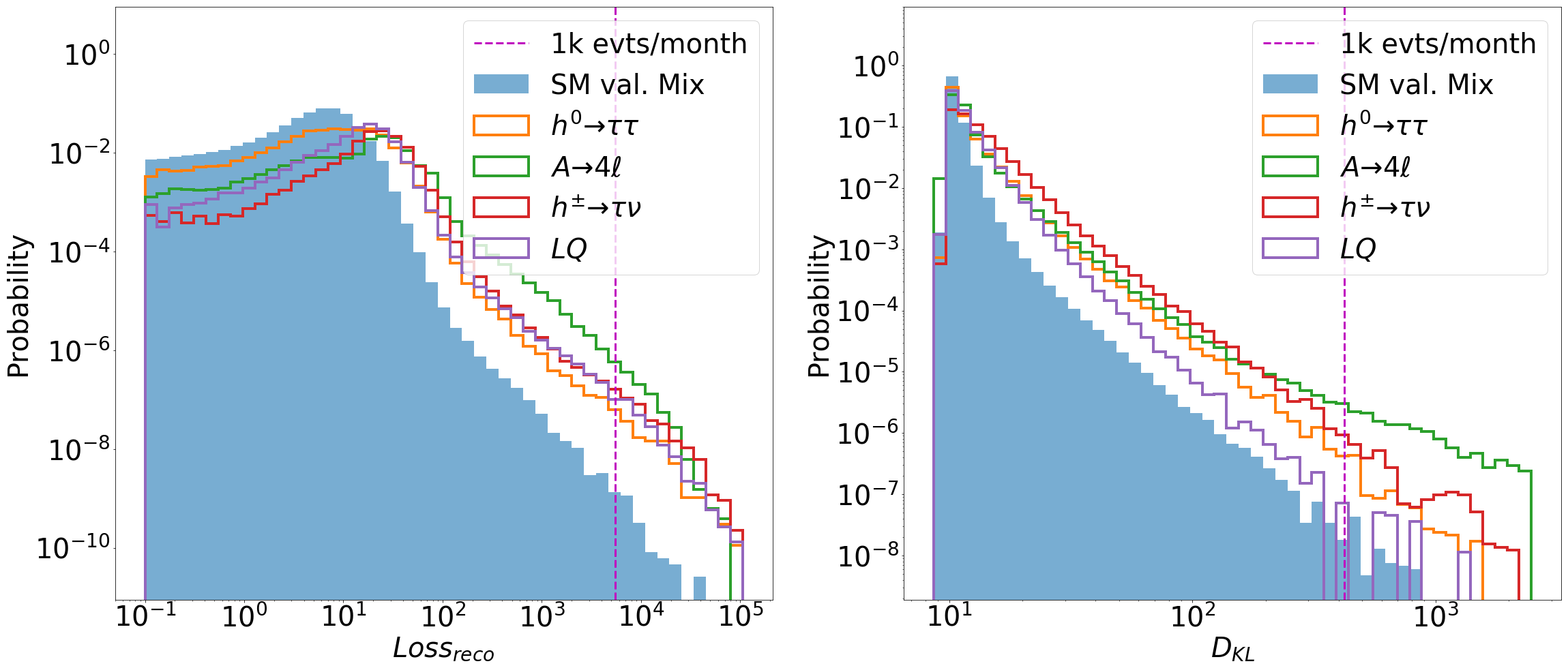}
    \caption{Distribution of the VAE's loss components, $\text{Loss}_\text{reco}$ (left) and $D_\text{KL}$ (right), for the validation dataset. For comparison, the corresponding distribution for the four benchmark BSM models are shown. The vertical line represents a lower threshold such that $5.4\cdot10^{-6}$ of the SM events would be retained, equivalent to $\sim 1000$ expected SM events per month.\label{fig:anomalytest}}
\end{figure}

Figure~\ref{fig:anomalytest} shows the distribution of the $\text{Loss}_\text{reco}$  and $D_\text{KL}$ loss components for the validation dataset. In both plots, the vertical line represents a lower threshold such that a fraction $\epsilon_{SM} = 5.4\cdot10^{-6}$ of the SM events would be retained. This threshold value would result in $\sim 1000$ SM events to be selected every month, i.e., a daily rate of $\sim 33$ SM events, as illustrated in Table~\ref{tab:VAE_results}. The acceptance rate is calculated assuming the LHC running conditions listed in Section~\ref{sec:intro}. Table~\ref{tab:VAE_results} also reports the by-process VAE selection efficiency and the relative background composition of the selected sample. 

Figure~\ref{fig:anomalytest} also shows the $\text{Loss}_\text{reco}$  and $D_\text{KL}$ distributions for the four benchmark BSM models. We observe that the discrimination power, loosely quantified by the integral of these distributions above threshold, is better for $\text{Loss}_\text{reco}$ than $D_\text{KL}$ and that the impact of the $D_\text{KL}$ term on $\text{Loss}_\text{Tot}$ is negligible. Anomalies are then defined as events laying on the right tail of the expected $\text{Loss}_\text{reco}$ distribution. Due to limited statistics in the training sample, the p-value corresponding to the chosen threshold value could be uncalibrated. This could result in a deviation of the observed rate from the expected value, an issue that one can address tuning the threshold. On the other hand, an uncalibrated p-value would also impact the number of collected BSM events, and the time needed to collect an appreciable amount of these events.

\begin{figure}[hp]
\centering
\includegraphics[width=0.85\textwidth]{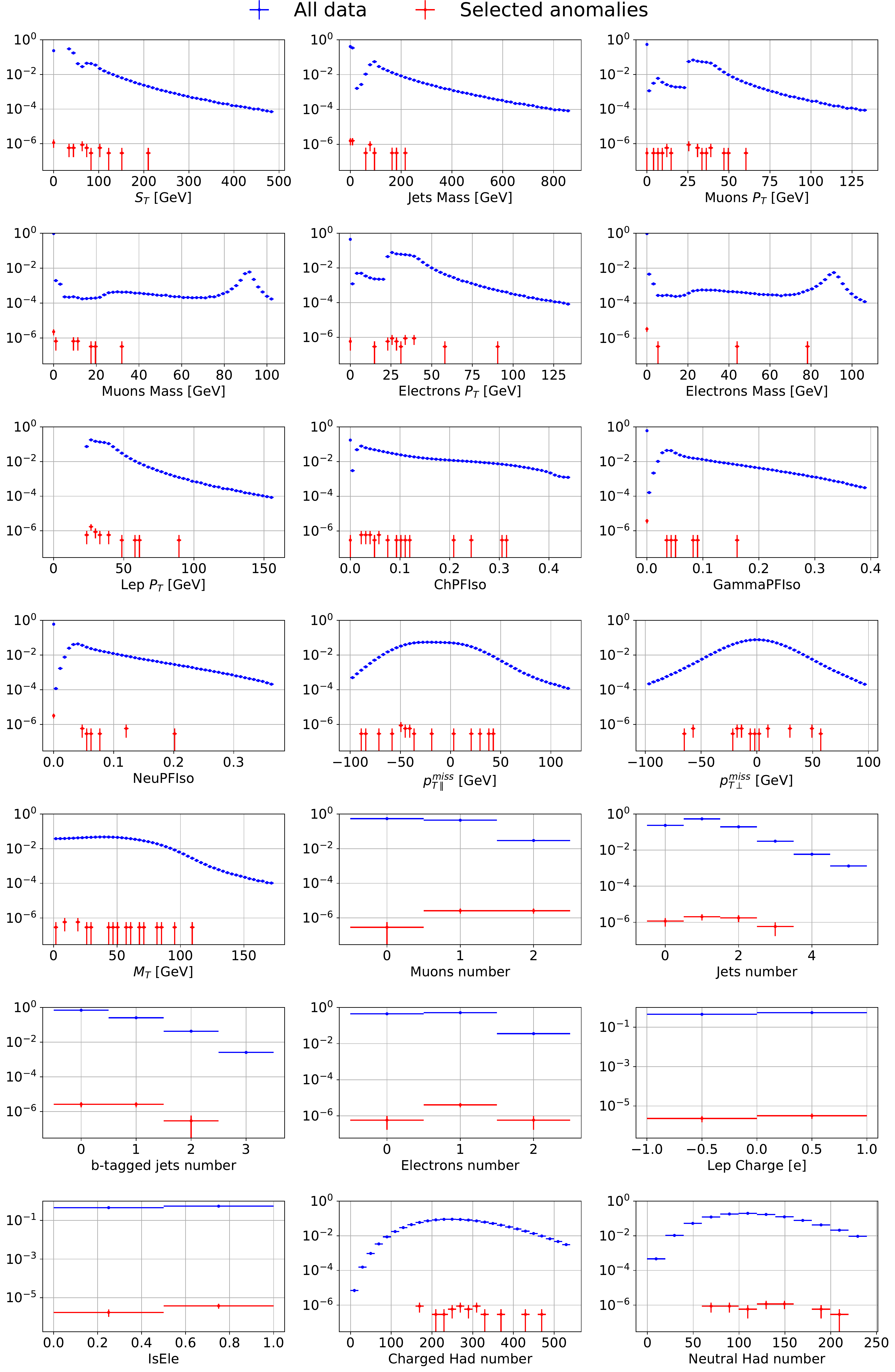}
\caption{Comparison between the input distribution for the 21 HLF of the validation dataset (blue histograms) and the distribution of the SM outlier events selected from the same sample by applying the $\text{Loss}_\text{reco}$ threshold (red dots). The outlier events cover a large portion of the HLF definition range and don't cluster on the tails.\label{fig:HLFdistrib_before_and_after}}
\end{figure}

\begin{figure}[hp]
\centering
\includegraphics[width=0.85\textwidth]{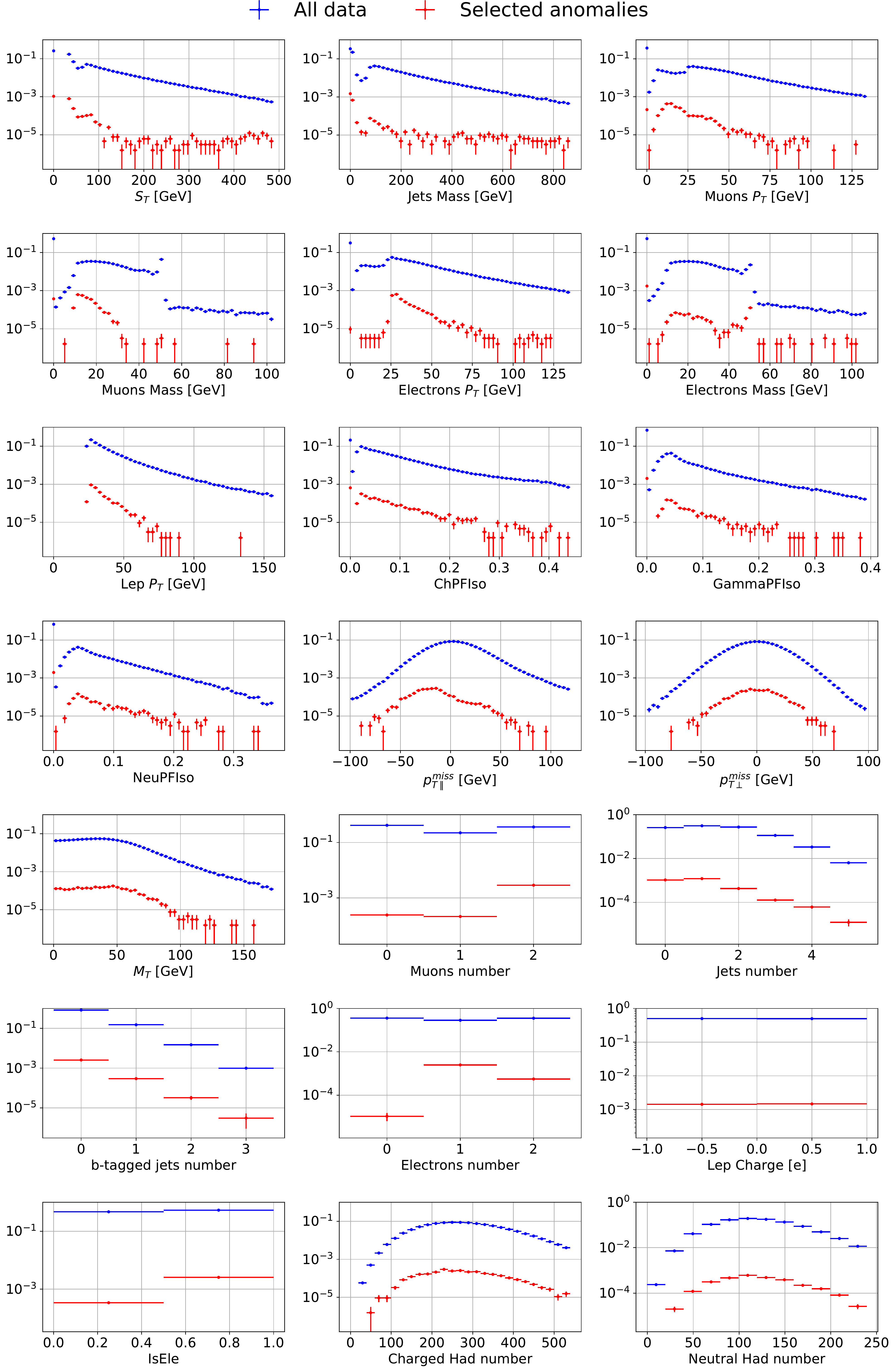}
\caption{Comparison between the distribution of the 21 HLF distribution for $A \to 4 \ell$ full dataset (blue) and $A \to 4 \ell$ events selected by applying the $\text{Loss}_\text{reco}$ threshold (red). 
The selected events are not trivially sampled from the tail.\label{fig:HLFdistrib_before_and_after_a4l}}
\end{figure}

Once the $\text{Loss}_\text{reco}$ selection is applied, the anomalous events don't cluster on the tails of the distributions of the input features. Instead, they tend to cover the full feature-definition range. This is an indication of the fact that the VAE does more than a simple selection of feature outliers, which is what is done by traditional single-lepton trigger or by dedicated cross triggers (e.g., triggers that select events with soft leptons and large missing transverse energy, $S_T$, etc.). This is shown in Fig.~\ref{fig:HLFdistrib_before_and_after} for SM events. A similar conclusion can be obtained from Fig.~\ref{fig:HLFdistrib_before_and_after_a4l}, showing the distribution of the 21 input HLF quantities for the $A \to 4\ell$ benchmark model, before and after applying the threshold requirement on the VAE loss. 

\begin{figure}[!htb]
\centering
\includegraphics[width=0.49\textwidth]{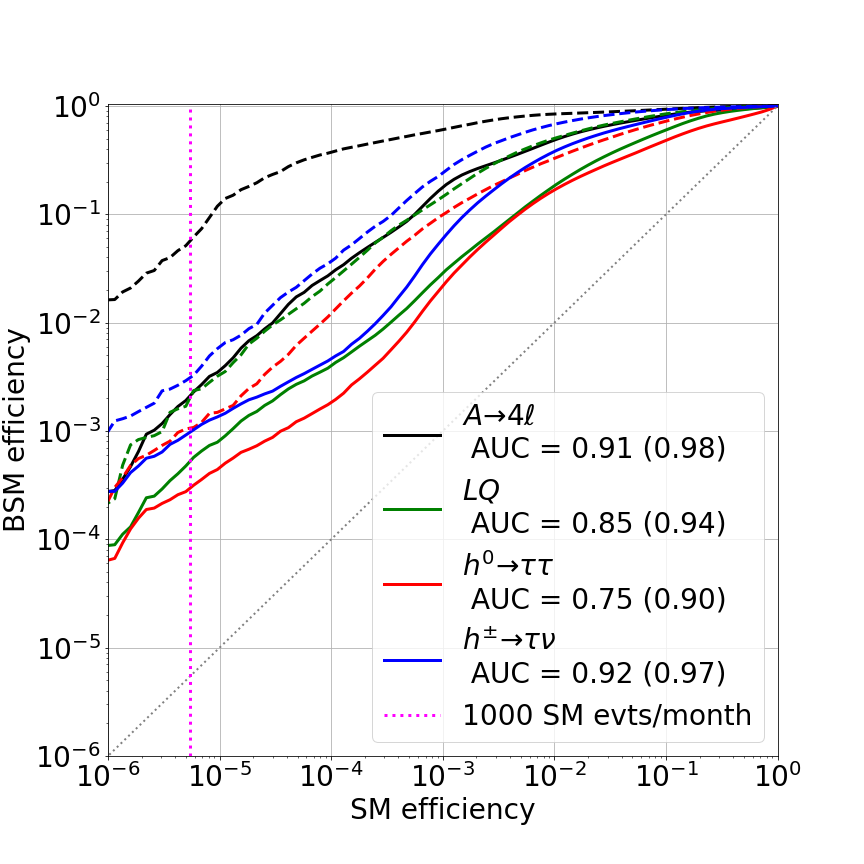}
\includegraphics[width=0.49\textwidth]{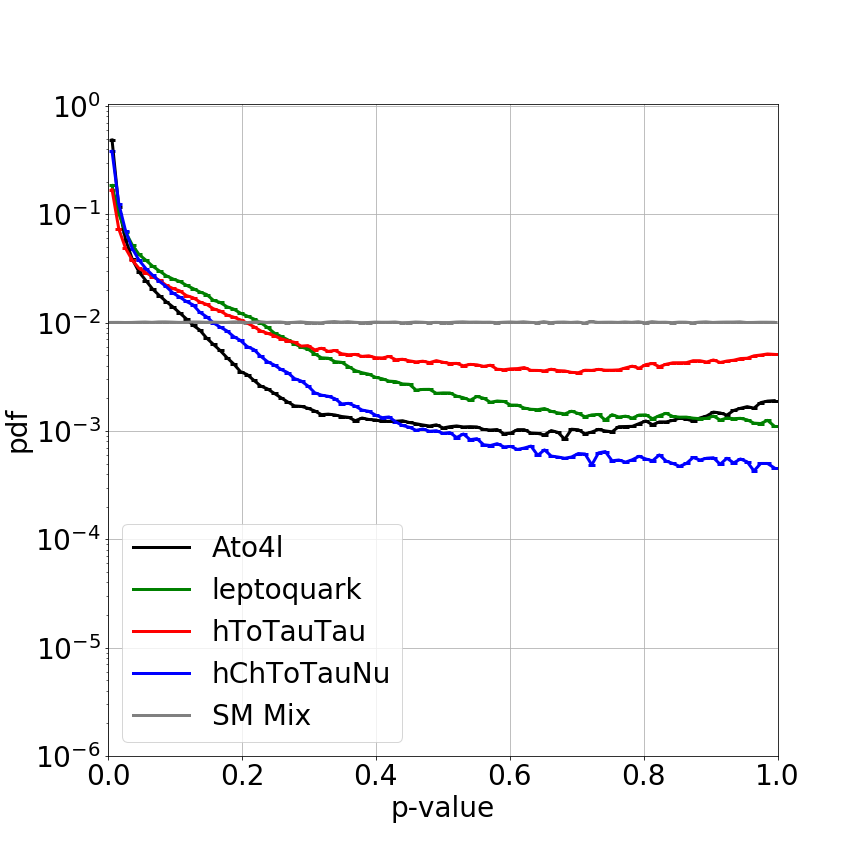}
\caption{Left:ROC curves for the VAE trained only on SM events (solid), compared to the corresponding curves for the four supervised BDT models (dashed) described in Section~\ref{sec:supervised}. Right: Normalized p-value distribution distribution for the SM cocktail events and the four BSM benchmark processes.\label{fig:models_ROC}}
\end{figure}

The left plot in Fig.~\ref{fig:models_ROC} shows the ROC curves obtained from the $\text{Loss}_\text{reco}$ distribution of the four BSM benchmark models and the SM cocktail, compared to the corresponding BDT curves of Section~\ref{sec:supervised}. As expected, the results obtained with the supervised BDTs outperform the VAE. On the other hand, the VAE can probe at the same time the four scenarios with comparable performances. This is a consequence of the trade off between precision and model independence and an illustration of the complementarity between the approach presented in this work and traditional supervised techniques. The right plot in Fig.~\ref{fig:models_ROC} shows the one-sided p-value computed from the cocktail SM distribution, both for the SM events themselves (flat by construction) and for the four BSM processes. As the plot shows, BSM processes tend to concentrate at small p-values, which allows their identification as anomalies. 

\begin{table}[!tb]
\caption{By-process acceptance rate for the anomaly detection algorithm described in the text, computed applying the threshold on  $\text{Loss}_\text{reco}$ shown in Fig.~\ref{fig:anomalytest}. The threshold is tuned such that a fraction of about $\epsilon_{SM} = 5.4\cdot 10^{-6}$ of SM events would be accepted, corresponding to $\sim 1000$ SM events/month, assuming the LHC running conditions listed in Section~\ref{sec:intro}. The sample composition refers to the subset of SM events accepted by the anomaly detection algorithm. All quoted uncertainties refer to $95\%$ CL regions.\label{tab:VAE_results}}

\vspace{10pt}

\centering
\begin{tabular}{c|c|c|c}
\hline
\multicolumn{4}{c}{Standard Model processes} \\
\hline
Process    & VAE selection                   & Sample composition    &  Events/month \\

\hline
$W$        & $3.6\pm 0.7 \cdot 10^{-6}$      & $32\%$              & $379\pm74$\\
QCD        & $6.0 \pm 2.3 \cdot 10^{-6}$    & $29\%$             & $357\pm143$\\
$Z$        & $21\pm3.5 \cdot 10^{-6}$      & $21\%$             & $256\pm43$\\
$t \bar t$ & $400 \pm 9\cdot 10^{-6}$      & $18\%$             & $212\pm5$\\
\hline
Tot        &                                 &                    & $1204\pm167$\\
\hline
\end{tabular}
\end{table}

Table~\ref{tab:VAE_results_BSM} summarizes the VAE's performance on the four BSM benchmark models. Together with the selection efficiency corresponding to $\epsilon_{SM} = 5.4\cdot 10^{-6}$, the table reports the effective cross section (cross section after applying the trigger requirements) that would correspond to 100 BSM events selected in a month (assuming an integrated luminosity of $5 \text{ fb}^{-1}$). Similarly, we quote the cross section that would result in a signal-to-background ratio of 1/3 on the sample of events selected by the VAE. The VAE can probe the four models down to small cross section values, comparable to the existing exclusion bounds for these mass ranges. As an example, Ref.~\cite{Chatrchyan:2012sv} excludes a $LQ \to \tau b$ with a mass of 150~GeV and production cross section larger than $\sim 10$~pb, using 4.8~fb$^{-1}$ at a center-of-mass energy of 7~TeV, while most recent searches~\cite{Sirunyan:2017yrk} cannot cover such a low mass value, due to trigger limitations.
\begin{table}[!htb]
\caption{Breakdown of BSM processes efficiency, and cross section values corresponding to 100 selected events in a month and to a signal-over-background ratio of 1/3 (i.e., an absolute yield of $\sim 400$ events/month). The monthly event yield is computed assuming an average luminosity per month of $5 \text{ fb}^{-1}$, computing by taking the LHC 2016 data delivery ($\sim 40\text{ fb}^{-1}$ collected in 8 months). All quoted efficiencies are computed fixing the VAE loss threshold $\epsilon_{SM} = 5.4\cdot 10^{-6}$.\label{tab:VAE_results_BSM}}

\vspace{10pt}

\centering
\begin{tabular}{c|c|c|c}
\hline
\multicolumn{4}{c}{BSM benchmark processes} \\
\hline
Process        & VAE selection                   & Cross-section         &  Cross-section \\
               & efficiency                      & 100 events/month [pb] &  S/B = 1/3 [pb] \\
\hline
$A \to 4\ell$  & $2.8 \cdot 10^{-3}$     & 7.1                   & 27 \\
$LQ\to b \tau$ & $6.7 \cdot 10^{-4}$     & 30                   & 110 \\
$h^0 \to \tau\tau$ & $3.6 \cdot 10^{-4}$     & 55                   & 210 \\
$h^\pm \to \tau\nu$ & $1.2 \cdot 10^{-3}$     &  17                  & 65 \\
\hline
\end{tabular}
\end{table}

Unlike a traditional trigger strategy, a VAE-based selection is mainly intended to select a high-purity sample of interesting event, at the cost of a typically small selection efficiency. To demonstrate this point, we consider a sample selected with the VAE and one selected using a typical inclusive single lepton trigger (SLT), consisting on a tighter selection than the one described in section~\ref{sec:data}. In particular, we require $p^{\ell}_T>27\text{ GeV}$ and $\text{ISO} < 0.25$. We consider the  signal-over-background ratio (SBR) for the VAE's threshold selection and the SLT. While these quantities depend on the production cross section of the considered BSM model, their ratio 
\begin{equation}
\frac{\text{SBR}_\text{VAE}}{\text{SBR}_\text{SLT}} = \left(\frac{\epsilon_\text{SLT}}{\epsilon_\text{VAE}}\right)_{SM} \cdot \left(\frac{\epsilon_\text{VAE}}{\epsilon_\text{SLT}}\right)_{BSM}
\end{equation}
is only a function of the selection efficiency for the SLT ($\epsilon_\text{SLT}$) and the for the VAE $\epsilon_\text{VAE}$ for SM and BSM events. Table~\ref{tab:SoB_improvement} shows how the SBR reached by the VAE is about two order of magnitude larger than what a traditional inclusive SLT could reach.

\begin{table}[!htb]
\caption{Selection efficiencies for a typical single lepton trigger (SLT) and the proposed VAE selection, shown for the four benchmark BSM models and for the SM cocktail. The last row quotes the corresponding BSM-to-SM ratio of  signal-over-background ratios (SBRs), quantifying the purity of the selected sample.\label{tab:SoB_improvement}}

\vspace{10pt}

\centering
\begin{tabular}{c|c|c|c|c|c}
\hline
 & SM & $A \to 4\ell$ & $LQ\to b \tau$ & $h^0 \to \tau\tau$ & $h^\pm \to \tau\nu$ \\
\hline
$\epsilon_\text{VAE}$ & $5.3 \cdot 10^{-6}$ & $2.8 \cdot 10^{-3}$ & $6.7 \cdot 10^{-4}$ & $3.6 \cdot 10^{-4}$ & $1.2 \cdot 10^{-3}$ \\
$\epsilon_\text{SLT}$ & 0.6 & 0.5 & 0.6 & 0.7 & 0.6 \\
$\epsilon_{SLT}/\epsilon_\text{VAE}$ & $1.1 \cdot 10^{5}$ & $1.8 \cdot 10^{2}$ & $9.0 \cdot 10^{2}$  & $1.7 \cdot 10^{3}$  & $5.8 \cdot 10^{2}$ \\
\hline
$\text{SBR}_\text{VAE}/\text{SBR}_\text{SLT}$ & - & 625 & 125 &70 & 191 \\
\hline
\end{tabular}
\end{table}

\section{How to deploy a VAE for BSM detection}
\label{sec:foreseen_applications}

The work presented in this paper suggests the possibility of deploying a VAE as a trigger algorithms associated to dedicated data streams. This trigger would isolate anomalous events, similarly to what was done by the CMS experiment at the beginning of the first LHC run. With early new physics signal being a possibility at the LHC start, the CMS experiment deployed online a set of algorithms (collectively called {\it hot line}) to select potentially interesting new-physics candidates. At that time, anomalies were characterized as events with high-$p_T$ particles or high particle multiplicities, in line with the kind of early-discovery new physics scenarios considered at that time. The events populating the hot-line stream were immediately processed at the CERN computing center (as opposed to traditional physics streams, that are processed after 48 hours). The hot-line algorithms were tuned to collect {\cal O}(10) events per day, which were then visually inspected by experts.

While the focus of the work presented in this paper is not an early discovery, the spirit of the application we propose would be similar: a set of VAEs deployed online would select a limited number of events every day. These events would be collected in a dedicated dataset and further analyzed. The analysis technique could go from visual inspection of the collisions to detailed studies of reconstructed objects, up to some kind of model-independent analysis of the collected dataset, e.g. a deep-learning implementation of a model-independent hypothesis testing~\cite{DAgnolo:2018cun} directly on the loss distribution (provided a reliable sample of background-only data).

While a pure SM sample to train VAEs could only be obtained from a MC simulation, the presence of outlier contamination in the training sample has typically a tiny impact on performance. One could then imagine to train the VAE models on so-far collected data and  use them on the events entering the HLT system. Such a training could happen offline on a dedicated dataset, e.g., deploying triggers randomly selecting events entering the last stage of the trigger system. The training could even happen online, assuming the availability of sufficient computing resources. As it happens with normal triggers, at the very beginning one would use some MC sample or some control sample from previously collected data to estimate the threshold corresponding to the target SM rate. Then, as it happens normally during HLT operations, the threshold will have to be monitored on real data and adjusted if needed.

\begin{figure}[!ht]
\centering
\includegraphics[width=0.6\textwidth]{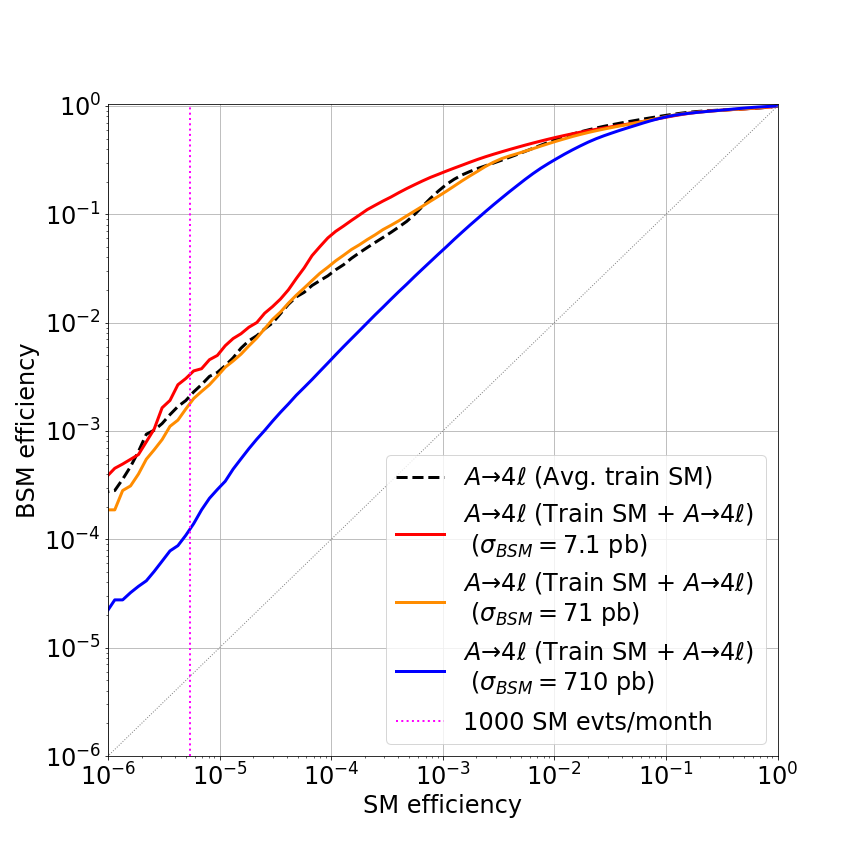}
\caption{ROC curves for the VAE trained on SM contaminated with and without $A\to4\mu$ contamination. Different levels of contamination are reported corresponding to $0.02\%$ ($\sigma = 7.15$ pb - equal to the estimated one to have 100 events per month), $0.19\%$ ($\sigma =  71.5$ pb) and $1.89\%$ ($\sigma =  715$ pb) of the training sample.\label{fig:ROC_sigCont}}
\end{figure}

To demonstrate the feasibility of a train-on-data strategy, we enrich the dataset used in  Section~\ref{sec:unsupervised} with a signal contamination of $A \to 4\ell$ events. As a starting point, the amount of injected signal is tuned to a luminosity of 100~pb$^{-1}$ and a cross section of 7.1~pb, corresponding to the value at which the VAE in Section~\ref{sec:unsupervised} would select 100 $A \to 4\ell$ events in one month. This results into about 700  $A \to 4\ell$ events added to the training sample. The VAE is trained following the procedure outlined in Section~\ref{sec:unsupervised} and its performance is compared to that obtained on a signal-free dataset of the same size. The comparison of the ROC curves for the two models is shown in Fig.~\ref{fig:ROC_sigCont}. In the same figure, we show similar results, derived injecting a $\times 10$ and $\times 100$ signal contamination. A performance degradation is observed once the signal cross section is set to 710~pb (i.e., 100 times larger than the sensitivity value found in Section~\ref{sec:unsupervised}). At that point, the contamination is so large that the signal becomes as abundant as $t \bar t$ events and would have easily detectable consequences. For comparison, at a production cross section of 27~pb a third of the events selected by the VAE in Section~\ref{sec:unsupervised} would come from  $A \to 4\ell$ production (see Table~\ref{tab:VAE_results_BSM}). Such a large yield would still have negligible consequences on the training quality.
This test shows that a robust anomaly-detecting VAE could be trained directly on data, even in presence of previously undetected (e.g., at Tevatron, 7~TeV and 8-TeV LHC) BSM signals.

The possibility of training the VAE on data would substantially simplify the implementation of the strategy proposed in this work, since any possible systematic bias in the data would be automatically taken into account during the training process. In addition, it would make the procedure robust against other systematic effects (e.g., energy scale, efficiency, etc.) that would affect a MC-based training.

\section{Conclusions and outlook}
\label{sec:conclusions}

We present a strategy to isolate potential BSM events produced by the LHC, using variational autoencoders trained on a reference SM sample. Such an algorithm could be used in the trigger system of  general-purpose LHC experiments to identify recurrent anomalies,  which might otherwise escape observation (e.g., being filtered out by a typical trigger selection). Taking as an example a single-lepton data stream, we show how such an algorithm could select datasets enriched with events originating from challenging BSM models. We also discuss how the algorithm could be trained directly on data, with no sizable performance loss, more robustness against systematic uncertainties, and a big simplification of the training and deployment procedure.

The main purpose of such an application is not to enhance the signal selection efficiency for BSM models. Indeed, this application is tuned to provide a high-purity sample of potentially interesting events. We showed that events produced by not-yet-excluded BSM models with cross sections in the range of ${\cal O}(10)$ to ${\cal O}(100)$ pb could be isolated in a $\sim 30\%$ pure sample of $\sim 43$ events selected per day. The price to pay to reach such a purity is a relatively small signal efficiency and a strong bias in the dataset definition, which makes these events marginal and difficult to use in a traditional data-driven and supervised search for new physics. 

The final outcome of this application would be a list of anomalous events, that the experimental collaborations could further scrutinize and even release as a catalog, similarly to what is typically done in other scientific domains. Repeated patterns in these events could motivate new scenarios for beyond-the-standard-model physics and inspire new searches, to be performed on future data with traditional supervised approaches.
 
We stress the fact that the power of the proposed approach is in its generality and not in its sensitivity to a particular BSM scenario. We show that a simple BDT could give a better discrimination capability for a given BSM hypothesis. On the other hand, such a supervised algorithm would not generalize to other BSM scenarios. The VAE, instead, comes with little model dependence and therefore generalizes to unforeseen BSM models. On the other hand, the VAE cannot guarantee an optimal performance in any scenario. As typical of autoencoders used for anomaly detection, our VAE model is trained to learn the  SM background at best, but there is no guarantee that the best SM-learning model will be the best anomaly detection algorithm. By definition, the anomaly detection capability of the algorithm does not enter the loss function, as well as, by construction, no signal event enters the training sample. This is the price to pay when trading discrimination power for model independence.

We believe that such an application could help extending the physics reach of the current and next stages of the CERN LHC. The proposed strategy is demonstrated for a single-lepton data stream coming from a typical L1 selection. On the other hand, this approach could be generalized to any other data stream coming from any L1 selection, so that the full $\sim 100$~Hz rate entering the HLT system of ATLAS or CMS could be scrutinized. 
 While  the L1 selection still represents a potentially dangerous bias, an algorithm running in the HLT could access 100 times more events than the $\sim 1$~kHz stream typically available for offline studies. Moreover, thanks to progresses in the deployment of deep neural networks on FPGA boards~\cite{Duarte:2018ite}, it is conceivable that VAEs for anomaly detection could be also deployed in the L1 trigger systems in a near future. In this way, the VAE would access the full  L1 input data stream.

\section*{Acknowledgments}

We thank D.~Rezende for his precious suggestions, which motivated us to explore Variational Autoencoders for this work. This project has received funding from the European Research Council (ERC) under the European Union's Horizon 2020 research and innovation program (grant agreement n$^o$ 772369) and the United States Department of Energy, Office of High Energy Physics Research under Caltech Contract No. DE-SC0011925. This work was conducted at  "\textit{iBanks}", the AI GPU cluster at Caltech. We acknowledge NVIDIA, SuperMicro  and the Kavli Foundation for their support of "\textit{iBanks}".

\clearpage
\appendix
\section{Comparison with Auto-Encoder}
\label{app:ae}
\begin{figure}[!htb]
\centering
\includegraphics[width=0.48\textwidth]{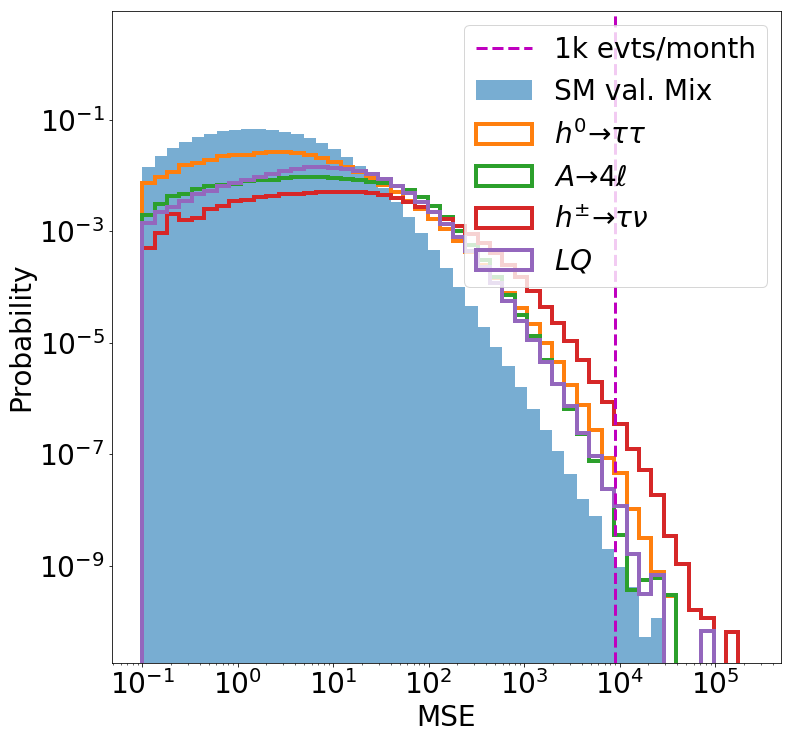}
\hspace{3pt}
\includegraphics[width=0.48\textwidth]{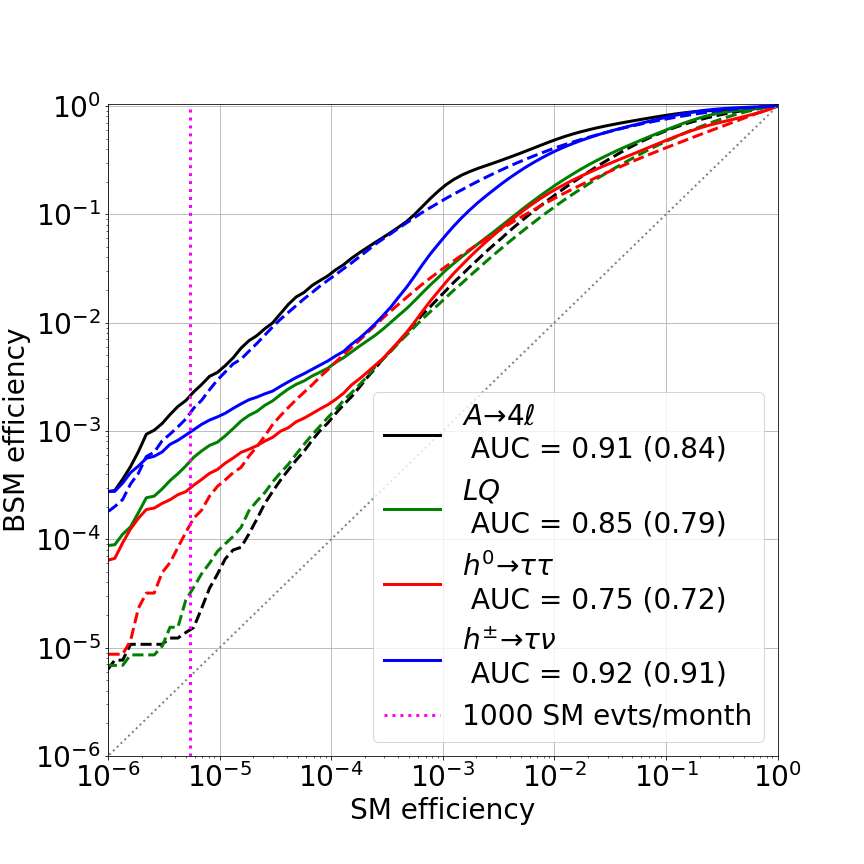}
\caption{
Left: Distribution of the AE loss (MSE) for the validation dataset. The distribution for the SM processes and the four benchmark BSM models are shown.
Right: ROC curves for the AE (dashed lines) trained only on SM mix, compared to the corresponding VAE curves from Fig.~\ref{fig:models_ROC} (solid). The vertical dotted line represents the $\epsilon_{SM} = 5.4\cdot 10^{-6}$ threshold considered in this study.\label{fig:AE_comparison}}
\end{figure}

For sake of completeness, we repeated the strategy presented in this work on a simple AE. The architecture was fixed to be as close as possible to that of the VAE introduced in Sec.~\ref{sec:unsupervised}. The change from VAE to AE imply these two changes: the output layer has the same dimensionality of the input layer; the latent layer includes four neurons (as opposed to 8), corresponding to the four latent variables $z$ (and not to the $\mu$ and $\sigma$ parameters of the $z$ distribution).  An MSE loss function is used. The optimizer and callbacks used to trained the VAE are are used in this case. Figure~\ref{fig:AE_comparison} shows the loss function distribution and a comparison between the ROC curves of the VAE and AE. These distributions directly compare to the left plots of Figs.~\ref{fig:anomalytest}~and~\ref{fig:models_ROC}, since in that case only the reconstruction part of the loss was used. For convenience, the VAE ROC curves are also shown here, represented by the dashed lines. 
When considering the four BSM benchmark models presented in this work, the AE provides competitive performances, for some choice of the SM accepted-event rate. On the other hand, the VAE usually outperforms a plain AE for the rate considered in this study ($\epsilon_{SM} = 5.4\cdot 10^{-6}$). With the exception of the $h^\pm \to \tau\nu$ model (for which the AE  provides a 30\% larger efficiency than the VAE), the VAE provides larger efficiency on the BSM models, with improvements as large as two orders of magnitude (for the $A \to 4\ell$  model).

\clearpage
\bibliographystyle{JHEP}
\bibliography{bib}

\end{document}